# Adiabatic Mode Transformation in Width-graded Nano-gratings: Enabling Multiwavelength Light Localization

Moein Shayegannia[1], Arthur O Montazeri [1,†], Katelyn Dixon[1], Rajiv Prinja[1], Nastaran Kazemi-Zanjani[1] and Nazir P Kherani[1,2,*]

[1]Department of Electrical and Computer Engineering, University of Toronto, Toronto, Ontario, M5S 3G4, Canada
[2]Department of Material Science & Engineering, University of Toronto, Toronto, Ontario, M5S 3E4, Canada
[†]Lawrence Berkeley National Laboratory, 1 Cyclotron Rd., Berkeley, CA, 94720, USA
*Email: kherani@ecf.utoronto.ca, Telephone: 416 946 7372, Fax: 416 946 8734
http://www.ecf.utoronto.ca/~kherani/

**Abstract:** We delineate the four principal surface plasmon polariton coupling and interaction mechanisms in subwavelength gratings, and demonstrate their significant roles in shaping the optical response of plasmonic gratings. Within the framework of width-graded metal-insulator-metal nano-gratings, mode confinement and wave guiding result in multiwavelength light localization provided conditions of adiabatic mode transformation are satisfied. The field is enhanced further through fine tuning of the groove-width (*w*), groove-depth (*L*) and groove-to-groove-separation (*d*). By juxtaposing the resonance modes of width-graded and non-graded gratings and defining the adiabaticity condition, we demonstrate the criticality of *w* and *d* in achieving adiabatic mode transformation among the grooves. We observe that the resonant wavelength of a graded grating corresponds to the properties of a single groove when the grooves are adiabatically coupled. We show that *L* plays an important function in defining the span of localized wavelengths. We show that multiwavelength resonant modes with intensity enhancement exceeding 3 orders of magnitude are possible with $w$<30nm and 300nm<$d$<900nm for a range of fixed values of *L*. This study presents a novel paradigm of deep-subwavelength adiabatically-coupled width-graded gratings – illustrating its versatility in design, hence its viability for applications ranging from surface enhanced Raman spectroscopy to multispectral imaging.

## Introduction

Spectroscopy – the study of light-matter interactions – is usually carried out through the dispersion of scattered light by diffraction gratings. High sensitivity spectroscopy of molecular species has been made possible by surface enhanced Raman scattering and surface enhanced fluorescence microscopy techniques [1-4]. These techniques benefit from resonant light-matter interactions owing to optimal light confinement within structures at nanometer length scales. Various nanostructures, such as nano-hole arrays [5, 6], gold nano-disks [7, 8], and metal-insulator-metal (MIM) waveguides [9, 10], localize the optical field within subwavelength cavities due to light coupling with its resonant modes. Among these nanostructures, MIM-based nano-gratings offer the distinct advantage of shaping its desired electromagnetic response, while simultaneously enhancing the optical intensity in both the near-field and far-field [9, 11-13].

To achieve enhanced localized fields within nano-gratings, optimal energy transfer is required between the incident light, surface plasmon polaritons (SPPs) propagating atop the grating structure, and the SPPs within the grooves. This can be achieved through impedance matching between the incident light and the SPPs (atop and within the grating) and adiabatic transformation of modes in between the grooves of a nano-grating [9, 14-16]. In general, adiabaticity is possible in graded nano-gratings or structures that support rainbow trapping [9, 16]. Geometric grading of nano-gratings can be either in depth or in width or both depth and width of the groove. Graded gratings provide the necessary multiwavelength enhancement required in certain applications such as single molecule detection in surface enhanced Raman spectroscopy, surface enhanced Fluorescence microscopy, or infrared spectroscopy [2]. Nevertheless, reducing the gap between the two adjacent metallic side-walls of the groove to the nanometer range within a grating intensifies electromagnetic fields within the cavity and in the near-field above it. Despite the challenges in fabrication of depth-based graded nano-gratings, width-based graded structures can be easily fabricated using planar lithography and thus open a new paradigm for plasmonic nano-gratings with a rich space of design parameters [2, 17, 18].

In this paper, we examine the significant role of surface plasmon polariton coupling and interaction mechanisms in subwavelength gratings *vis-à-vis* shaping the optical response of the grating. We undertake an in-depth parametric investigation of the requisite conditions for adiabaticity in width graded nano-gratings. Specifically, we carry out controlled perturbative variations in depth ($L$), width ($w$), and groove-to-groove separation ($d$) to determine the parameter set of $L$, $w$ and $d$ that yield adiabatic mode transformation and thus impedance matching within the width graded nano-gratings. We further contrast resonant modes in graded and non-graded gratings and thus provide insight into the cruciality of the geometric parameters $w$ and $d$ in attaining adiabaticity in graded nano-gratings as well as in shaping the electric field profile (intensity, FWHM, and $Q$-factor).

Here below, following the introduction of the concept of resonance in periodic and graded MIM gratings we delineate the four principal coupling mechanisms that define the resonant condition in graded gratings. Analytically we develop the conditions for adiabaticity in width-graded gratings. Then we present detailed parametric studies of:

I. Single nano-groove; II. Non-graded/Uniform nano-gratings; and III. Width-graded nano-gratings.

## Results

By availing the width of nano-grooves as a tunable parameter, we open a new avenue in light trapping within the field of subwavelength chirped gratings. Groove widths of less than 30 nm become a principle geometric parameter for effective light trapping in MIM gratings. This is due to the coupling of the evanescent plasmonic fields on the side walls of the groove. The schematic of the width-graded nano-grating under study in this article is shown in Figure 1.a where the smallest groove is situated at the center, and is surrounded by nano-grooves with increasing groove widths on either side. Figure 1.b shows a cartoon of multiwavelength light trapping within the width-graded grating, achieved through parametric optimization of phase engineering that results in synergistic SPP trapping and waveguiding effects – the details of which are described in the following sections. For brevity, hereafter we refer to the incident wavelength simply by $\lambda$ and denote the two types of lamellar MIM gratings as either 'uniform gratings' or 'graded gratings'.

### Concept of Resonant Condition

To provide strong mode confinement within the nanogrooves of periodic and graded MIM gratings, two conditions must be satisfied [19]. The first, a phase matching condition, which is:

$$\beta_j - \beta_{j\pm 1} = m\left(\frac{2\pi}{\Lambda_j} - \frac{2\pi}{\Lambda_{j\pm 1}}\right) , \quad (1)$$

where $\beta_j$ is the plasmonic propagation mode constant in the $j^{th}$ groove, $m$ is the mode number, and $\Lambda$ is the grating period. In a graded grating, the period increases with groove width ($w_j = w + j\Delta$), that is, $\Lambda_j = d + w_j$. The second condition is the coupling between field amplitudes of the resonant modes slowly moving in adjacent grooves. This is given by [19]:

$$dA_j = -i\frac{\beta_j}{|\beta_j|}C^m_{jj\pm1}A_{j+1(x)}e^{i\left(\beta_j-\beta_{j+1}-\frac{m2\pi}{\Lambda_j}\right)x}dx \quad (2)$$

where $A_j$ represents amplitude of the $m^{th}$ resonant mode in the $j^{th}$ groove, and $C^m_{jj\pm1}$ is the magnitude of coupling between the $m^{th}$ mode in the $j^{th}$ and ($j$+1$^{th}$ groove, which is defined as:

$$C^m_{jj\pm1} = \frac{\pi}{2\lambda}\int E_j^*.\varepsilon_m(x)E_{j\pm1}dx\ , \quad (3)$$

where $E_j$ is the localized electric field in the $j^{th}$ groove, and $\varepsilon_m(x)$ is the dielectric perturbation; for a periodic grating $\varepsilon_m$ is independent of $x$ while for a graded grating $\varepsilon_m$ is a function of $x$.

To satisfy the phase matching condition of Eq. (1) and to attain a strong coupling coefficient per Eqs. (2) and (3), four primary coupling mechanisms are identified between the incident field and the SPPs on the surface and on the sidewalls of a subwavelength grating which contribute to the resonant behavior of the structure. Each of these coupling mechanisms is illustrated in Figure 1.c; for brevity, we show each of these in a single groove yet recognize that any groove might support any number of these mechanisms to localize a given mode.

**Coupling Mechanisms**

I. *Coupling of incident wave to the cavity mode*

An incident plane wave couples to a grating structure through double reflection Fabry-Perot resonator modes. Specifically, the impedance mismatch between the effective refractive index of each groove (local mode index) and free space causes reflection of the coupled wave at the top of the grating. A similar reflection of the wave also occurs at the bottom (metallic surface) of the groove. These give rise to multiple reflections and thus a standing wave [20]. For a very small groove width ($w$), only the fundamental mode plays a significant role, while all other modes are strongly evanescent in the groove [21]. This coupling mechanism is illustrated in groove $j$ =1 in Figure 1.c, and can be described by [22]:

$$4n_{\mathrm{eff}}L\cos\theta = (2m-1)\lambda_0, \tag{4}$$

where $n_{eff}$ is the effective index of the groove, $\theta$ is the angle of incidence, $L$ is the groove depth, $m$ is the mode number, and $\lambda_0$ is the incident wavelength.

II. *Coupling of surface waves to a cavity mode*

Surface electromagnetic waves or surface plasmon polaritons are transverse magnetic (TM) traveling waves along an interface between two media with opposite signs of the real part of the dielectric constant. A groove carved into a metal film creates a local optical inhomogeneity which makes possible the matching of the in-plane $k$-vectors of the incident wave and the SPPs. The proximal repetition of such grooves—forming a subwavelength grating—guides the incident light along the surface, while forming surface plasmon polaritons. Conservation of momentum dictates that the coupling between the wave vector of light and the wave vector of surface plasmon polaritons occurs with no change in frequency, albeit the wavelength can be compressed by changes in the effective index [19]. SPPs, at the interface of the grating surface and air, travel down the side-walls of the grooves—coupling with the fundamental photonic mode of that groove—and lead to high electric field intensity inside the grooves. This coupling mechanism is illustrated in groove $j = 3$ of Figure 1.c and is described by [20]:

$$\beta_j = k \sin\theta \pm m\frac{2\pi}{\Lambda_j}, \tag{5}$$

*III. Coupling between the SPP modes on the side-walls of an MIM cavity*

For very narrow groove widths ($w$ ≲100 nm) in MIM structures, each groove serves as a modified Fabry-Perot resonator [23]. The SPPs excited on the side-walls of a narrow subwavelength groove, where $w << \lambda$ and the effective index is high, couple by virtue of the overlapping evanescent fields. This coupling leads to a significant electric field enhancement within the narrow groove [14]. The effective index is calculated using the following set of equations [20]:

$$\tanh k_1 \frac{w}{2} = -\frac{k_2\varepsilon_1}{k_1\varepsilon_2}, \tag{6}$$

$$k_i^2 = \beta_j^2 - k_0^2\varepsilon_i, i = 1{,}2, \tag{7}$$

$$\beta_j = n_{\mathrm{eff}_j} k_0, \tag{8}$$

where $k_1$ and $k_2$ represent the wavevector of light in the metal and insulator in the MIM cavities, respectively. In contrast, in the limit of wide grooves ($w \approx \lambda$ or $w \geq 100$ nm) [23], the dispersion relation of the SPPs on the side-walls of the groove is similar to that of a semi-infinite dielectric-metal interface [20]. These coupling mechanisms are illustrated in grooves $j$ = 0 (intra-groove coupling of side-wall SPPs) and $j$ = 3 (weak or no coupling of side-wall SPPs), respectively, in Figure 1.c.

The theoretical dependence of the resonant wavelength of a single groove on its width ($w$) and depth ($L$), as described by Eqs. (4) to (8), is illustrated in Figure 2.

*IV. Coupling between evanescent modes of adjacent grooves*

In grooves $j$ = -2, -1, and 0 of Figure 1.c, a localized groove mode, or a confined SPP within the groove, penetrates the inter-groove metal and thus couples to the adjacent groove modes. The inter-groove penetration of these modes is caused by the finite conductivity of the metal in the visible and near infrared region, and it occurs if the penetration depth of these evanescent SPP modes ($\delta_{spp}$) is comparable to the separation between the grooves ($d$) [15]. Two asymptotic regimes are considered in this case: the optical regime (near/mid infrared) and the electrostatic regime (visible wavelength). In the optical regime, $\delta_{spp} \approx \delta_{metal}$, whereas in the electrostatic region, $\delta_{spp} << \delta_{metal}$, where $\delta_{metal}$ is the skin depth of the metal [15]. Groove width and excitation wavelength determine the crossover point between these two regimes. These parameters and the separation between adjacent grooves determine whether an evanescent mode from one groove can couple to that in an adjacent groove.

## Concept of Adiabatic Mode Transformation

To achieve multiwavelength localization in a graded grating, perturbation in the dielectric properties needs to occur sufficiently slowly so as to allow an adiabatic process whereby dissipation of the EM energy is minimized. Under adiabatic mode transformation, the graded grating is impedance-matched; that is, each groove resonates strongly at a single wavelength while transferring non-resonant modes to an adjacent groove(s). (for example, see the cartoon in Figure 1.b, and Figures 5, 6). The adiabatic parameter $\delta$ is derived using the Wentzel-Kramers-Brillouin approximation [24] and by simplifying Schrodinger's equation for a one-dimensional motion of a single particle in a "quasi-classical" system. This approximation necessitates [25]:

$$\left|\frac{\hbar\sigma''(x)}{(\sigma'(x))^2}\right| \ll 1 \tag{9}$$

where ħ is the reduced planks constant, $\sigma'(x)$ and $\sigma''(x)$ are the first and second derivatives of the $\sigma(x)$, the phase in the solution of Schrodinger's equation wave function approximated by $\sigma(x)\approx\int p(x)dx$, where $p(x)$ is the momentum of a particle with a mass of $m$, total energy $E$, and potential energy $V$(x). The momentum of a particle is defined as $p(x) = \sqrt{2m[E - V(x)]} = \hbar\beta$ [25]. So the approximation in Eq. (9) reduces to the following equation which is the adiabatic parameter for the $j^{\text{th}}$ groove:

$$\delta_j = \frac{\frac{1}{\beta_j} - \frac{1}{\beta_{j+1}}}{\Lambda_j} \tag{10}$$

where $\Lambda_j = d + w_j$ and $w_j = w + j\Delta$ as shown in Figure 1.a . Figure 3 shows the adiabatic parameter (calculated on the grating surface) as a function of the period and the gradient in groove width. It is observed that larger values of the period and smaller values of the adiabatic parameter $\delta$ improve the adiabaticity between the grooves.

**Single groove (n = 1)**

In this section, we investigate the impact of the groove width ($w$) on the resonance behaviour of a single groove carved in a metallic slab. The local optical inhomogeneity created by the single groove is not sufficient to match the in-plane *k*-vectors of the incident light with the SPPs. Thus, the field localization within a single groove seen in Figure 4 is mostly driven by the photonic modes inside the cavity that directly couple to the incident light through an end-fire coupling mechanism [26]. Also, Figure 4 shows that stepwise increase in *w* from 10 nm to 20 nm correspondingly decreases the resonance wavelength ($\lambda_{resonant}$) from ~5.2 μm to ~3.5 μm in the mid-infrared regime, or ~1600 nm to ~1000 nm in the near-infrared regime.

Close examination of Figure 4 indicates a counterintuitive inverse relationship between *w* and $\lambda_{resonant}$, in contrast to that in a Fabry-Perot cavity where larger grooves support longer plasmonic wavelengths; that is, the plasmonic resonant wavelength within the groove increases with decreasing groove width (*w*). The underlying reason for this behaviour is the fact that the effective index of the groove increases as it narrows [9]. Hence, for a fixed groove depth the wavelength of the Fabry-Perot mode as supported by a narrower groove increases in accordance to Eq. (4).

At resonance, the incident wave, SPPs, and other surface waves on the metal surface couple to the cavity mode leading to a high intensity electric field localized inside the groove. The localized surface plasmons reradiate into free space and dissipate in the surrounding metal within the cavity. However, at wavelengths below or above the resonance, the surface waves do not couple to the cavity modes and propagate away from the groove. Two regimes of operation can be realized based on the groove width in a grating structure: electrostatic regime (for $w < 10$ nm) and optical regime (for $w > 10$ nm) [15]. For visible excitation frequencies, decreasing the width of the groove down to a few nanometers moves the dispersion of the guided mode within the grooves from the optical to the electrostatic regime. For example, at $\lambda = 500$ nm, the crossover point between these two regimes is a groove width of 10 nm [15]. Whereas at larger wavelengths in the infrared, the crossover point is at a smaller groove width. To carry out consistent numerical analyses, we performed the simulation in the optical regime where the minimum groove width is kept at *w* = 10 nm.

Our simulation results also indicate that reducing the groove width reduces the FWHM of the resonant peak of the groove and thus enhances the *Q*-factor. The *Q*-factor is calculated using $\nu_{\text{resonant}}/\text{FWHM}$, where $\nu_{\text{resonant}}$ is the resonant frequency of the groove. In the mid-infrared regime, for simulated groove widths of *w* = 10, 15, and 20 nm, FWHM (*Q*-factor) assume values of 2.9 μm (1.6), 1.5 μm (2.3), and 0.7 μm (5.5), respectively. In the visible regime, these values change to 35 nm (19.5), 36 nm (19.4), and 37 nm (17.5) for groove widths of *w* = 10, 15, and 20 nm, respectively. It is observed that in the mid-infrared regime, FWHM (*Q* factor) has a decreasing (increasing) trend with increasing groove width, while in the visible regime this trend changes to increasing (decreasing). Although, the *Q* factors appear to be small, *Q* factor values normalized to the volume (*Q*/*V*) for such subwavelength gratings are comparable to those of conventional optical resonators [20]. Please see the supplementary information for a comparison of the *Q/V* values between our graded grating and those of optical resonators.

The theoretical data in Figure 2 and the simulation results in Figure 4 indicate that multiwavelength localization is possible via proximal repetition of subwavelength nano-grooves that have a gradient in their widths.

**Multiple grooves with constant groove width (n > 1 and Δ = 0)**

We next simulate a uniform lamellar grating with a varying number of grooves to investigate the resonance profile, localized field intensity and FWHM of the grooves. The results are summarized in Table 1.

Proximal repetition of nano-grooves reinforces SPPs over the grating surface. Increasing the number of grooves ($n$) introduces the groove-to-groove separation ($d$) as a new degree of freedom and accordingly improves matching of the in-plane $k$-vectors of an incident wave and the SPPs. This strengthens SPPs on the top interface between the grating and air. When $d$ = 5 nm, the separation between the grooves is less than the skin depth of the metal ($\delta_{metal}$), consequently the evanescent mode-coupling between the adjacent grooves (inter-groove coupling) plays a significant role in defining the resonance profile of the structure. In this case, the resonant cavity modes tunnel through the metal, without necessarily coupling to the SPPs on the grating surface. This results in a lower intensity of the localized electric field within the grooves which in turn lowers the effective $Q$-factor (i.e., increases the FWHM), where the effective $Q$-factor of a grating structure is taken as the average of the individual $Q$-factors of the grooves within the given grating structure.

The SPPs propagating on the grating surface and other non-propagative SPPs resulting from the scattering of light at the groove edges can couple to the Fabry-Perot modes inside the grooves. For $d > \delta_{metal}$, the constructive interference between the SPPs on the grating surface and their coupling to the Fabry-Perot modes define the resonance profile of the structure. For example, at $d$ = 300 nm and 700 nm such constructive interference leads to a high intensity of the localized field. Upon increasing $d$ further, the SPPs propagate farther on the metallic surface of the grating prior to reaching the next groove and accordingly they experience a decay in intensity due to the ohmic losses. For example, at $d$ = 1000 nm the Fabry-Perot modes inside the grooves only directly couple to the incident light. Table 1 shows that the latter coupling mechanism not only leads to a relatively lower localized electric field intensity but also a smaller effective $Q$-factor.

Identical groove depth, width, and groove-to-groove separation in a uniform grating implies that such a structure would only resonate at a single wavelength of light. However, Table 1 shows that this structure resonates at two different wavelengths at a groove depth of $L$ = 100nm, and higher order modes are excited at deeper groove depth of $L$ = 500nm, as shown in Figure 5.a. According to Eq. (5), the propagation constant of the SPPs coupling to the cavity modes (mechanism II) can be both positive or negative depending on its direction of propagation, which leads to localization of two different wavelengths.

As $d$ increases, the SPPs on the grating surface decay and the presence of bi-wavelength localization reduces to uni-wavelength localization in uniform gratings. For example, at $d$ = 1000 nm and $n$ = 3 (see Table 1), all three grooves localize a single wavelength at $\lambda$ = 1166 nm, whereas at $d$ = 300 nm bi-wavelength localization is achieved throughout the grating structure. Evidently, localization of more than one wavelength of light in a nano-grating structure is linked to the strength of the SPPs propagating atop the grating surface and their coupling to the cavity modes where available.

As $n$ increases to infinity, diffraction orders in a uniform lamellar grating begin to emerge for $d \geq \lambda$. Starting from $d \approx 600$ nm, every groove acts as a quasi-point source and higher order diffraction modes emerge. Described by the Huygens−Fresnel principle, the propagating diffracted wave is the superposition of the secondary wavelets created by these point sources. As a result, the structure is also a diffraction grating and the diffraction orders add to the propagation constant of the SPPs according to Eq. (4). The magnitude of the first order diffracted wave is at its maximum at $d$ = 700 nm. This effect gives rise to a very small FWHM of 1 nm and a very high $Q$-factor of 723 (i.e., at $n = \infty$, $d$ = 700 nm and $\lambda_{\text{resonant}} = 723$ nm (see Table 1)). At larger values of $d$, with $n$ still at infinity, higher order diffracted modes also contribute, albeit insignificantly to the propagation constant of the SPPs and lead to low intensity higher order peaks which have larger FWHMs.

**Graded Gratings ($\Delta \neq 0$)**

Next, we carry out a series of simulations to study the localized electric field profiles in the graded gratings. In these simulations, we set $w$ = 10 nm and $\Delta$ = 5 nm, to stay within the optical regime, as mentioned in the section on single groove. To simplify the simulations, we considered a non-symmetric (one-sided) graded grating with $n$ = 5 and $j$ ranging from 0 to -4, as shown in Figure 1.a. The value of $L$ and $d$ are varied from 10 nm to 1 µm, and 5 nm to 10 µm, respectively; we present the results for $L$ = 500 nm and $d$ = 300 nm (see Figure 5.b and Figure 6) considering that adiabatic coupling is improved when 300 nm $\leq$ d $\leq$ 900 nm (which is discussed further below).

In a graded grating structure, the effective refractive index of the grooves increases with decreasing groove width; that is, a width-graded grating is also a structure graded in effective index. Thus, the group velocity of the SPPs traveling on the grating surface changes with the width of the groove and reaches a minimum across the narrowest groove. This variation in group velocity along with the constraint of conservation of energy gives rise to multiwavelength light localization in a width-graded grating as defined here (where $d$ is a constant).

The multiwavelength localization shown in Figure 5.b is of an impedance-matched graded grating which benefits from adiabatic mode transformation. Such graded grating structures provide multiwavelength localization of light over a large bandwidth that stretches from approximately 600 nm to about 6 μm, in contrast to a uniform grating (Figure 5.a) which offers only uni-wavelength localization in discrete spectrum range.

Figure 3 demonstrates that larger values of the period lead to better adiabaticity between the grooves. Nevertheless, as noted from Table 1, intensity of the SPPs on the grating surface decay for $d$ > 1μm, due to ohmic losses. As a result, at 300 nm ≤ $d$ ≤ 900 nm improved phase engineering of the SPPs leads to adiabatic mode transformation between adjacent grooves and thus leads to enhanced multiwavelength localization in the graded gratings (as shown in Figure 5.b for $d$ = 300 nm). Under adiabatic mode transformation, the graded grating is impedance-matched, that is, each groove resonates strongly at a single wavelength while transferring their non-resonant modes to an adjacent groove(s) so that the modes overlap with the resonant mode of the neighbouring groove(s).

Figure 6 shows how adjacent grooves couple and guide their non-resonant modes to the neighbouring resonating mode in an impedance-matched graded grating. The blue vectors pointing to the right (left) correspond to positive (negative) values of the $x$-component of the electric field ($\vec{E}_x$). The resonating groove, indicated on the figure, exhibits the highest field strength ($E = \left(E_x^2 + E_y^2 + E_z^2\right)^{1/2}$) which evidently corresponds to the highest total power dissipated at the bottom of the groove. $\vec{E}_x$ of the resonating groove points to the right while the field in all the other grooves point toward the resonating groove. For example, when groove $j = 0$ (Figure 6.a) is at resonance (at $\lambda_{resonant} = 5.264$ μm), $\vec{E}_x$ of all the other grooves point towards it. Similarly, when groove $j = -1$ is at resonance (at $\lambda_{resonant} = 4.478$ μm), $\vec{E}_x$ of all the other grooves point toward the resonating groove. Our simulation results show that this adiabatic mode transformation occurs for every groove only when 300 nm ≤ $d$ ≤ 900 nm. This in turn leads to enhanced multiwavelength localization of light. Under these conditions of adiabaticity, the graded grating structure is said to be impedance matched such that the energy is efficiently transferred from the incident light to the SPPs on the surface of the grating and to the photonic modes of the grooves. For 50 nm ≤ $d$ < 300 nm, $\vec{E}_x$ in the grating changes its direction multiple times above the resonant frequency and thus does not allow coupling of a single dominant plasmonic resonant mode to the grooves, thereby leading to poor multiwavelength localization. (For example, see supplementary information Visualization 1 animating the behavior of $\vec{E}_x$ and the resonance profile of the graded grating for $d$ = 100 nm, and $d$ = 300 nm, at $L$ = 100 nm.)

The reason for non-adiabatic behaviour of the $\vec{E}_x$ for $d$ < 300 nm is that the propagation constant of the SPPs on the grating surface is smaller than those resonating within the groove, thus preventing their coupling. However, under adiabatic mode transformation, for 300 nm ≤ $d$ ≤ 900 nm, the propagation constant of the SPPs on the grating surface is larger or equal to those resonating within the groove which allows for mode coupling in between the grooves, based on Eqs. (6) and (8).

Our simulation results also reveal that contrary to the redshifting of the resonant wavelength in uniform gratings with increasing $d$, in the graded grating the resonant wavelength of the grooves neither redshifts nor blueshifts with changing $d$. Hence, the resonant wavelength(s) of each groove in an adiabatically coupled graded grating structure is simply determined by examining the resonant wavelength(s) of the corresponding single groove. Inspecting Figures 4, 5.b, and Table 2, we observe that identical resonant wavelengths are realized for grooves with identical groove width in both the single groove structure and the graded grating structure. However, determination of the intensity and FWHM for a resonant mode in a graded grating requires modeling of the grating structure as a whole. These results corroborate a previous study wherein a simple model for a single groove was used to explain the behavior of an entire functionally graded grating with non-homogenous effective index [9].

Table 2 shows the resonance profile of a (non-symmetric) one-sided graded grating. The resonance wavelengths and intensities of this structure are similar to that of a symmetrically graded grating, except for the localized field intensity in the center groove ($j$ = 0) of the symmetric grating which is twice as large as that in the non-symmetric grating. This fact demonstrates that the graded grating not only localizes multiwavelength resonant modes, but also serves as a waveguide to propagate the off-resonant modes from the outer-lying wider grooves toward the narrower central-lying grooves.

FWHM and $Q$ factor of the graded gratings for different values of $d$ is tabulated in Table 2. An optimized graded grating structure should possess a small FWHM, large $Q$ factor, and multiwavelength localization in all grooves. Based on Table 2, the FWHM and Q factor of a shallow graded grating with $L$ = 100 nm ranges between 146 nm to 185 nm and 5.28 to 8.57, respectively. Whereas a deeper graded grating with $L$ = 500 nm offers better values of FWHM and Q factor for higher order modes, albeit at the cost of reduced intensity for these localized modes. Our simulation results also indicate that using other dielectrics such as $SiO_2$ instead of air in the MIM grooves introduces a higher index contrast and lowers the average FWHM at the cost of the localized intensity. For example, using $SiO_2$ as an insulator in the MIM grooves at $L$ = 100 nm yields a FWHM of 169 nm, an effective $Q$ factor of 9.2, and an intensity of $0.87 \times 10^{10}$ ($V^2/m^2$) for $j$ = -1.

A graded grating structure with the resonant profile shown in Figure 5.a offers enhanced rainbow trapping in the visible, near-infrared, and mid-infrared spectrum range with many potential benefits for surface enhanced Raman, fluorescence, and infrared spectroscopy, and hence sensing applications, by virtue of overlapping vibrational modes of interest [1-3], [27, 28]. Further, fabrication of such high aspect ratio structures is feasible via precise lithographic techniques [29], facile multi-layer sputter-deposition of thin metal-dielectric layers [18], and simple volume manufacturing through nano-imprint lithography and peel of techniques [refs].

## Conclusion

We have identified four different coupling mechanisms that underpin the plasmonic resonance profile of sub-wavelength uniform and graded lamellar gratings in the visible and near-mid infrared regions. We have utilized this framework to develop unique insight into the underlying physics of adiabatic mode transformation between the nanogrooves which leads to multiwavelength light localization in a graded grating. The simulation results for uniform and graded gratings reveal that evanescent mode-coupling of adjacent grooves (with groove-to-groove separations of less than 50 nm) lowers the $Q$ factor of the structure which in turn leads to a diffusion of the resonance profiles over all the grooves—groove hybridization—and hence impairs multiwavelength localization. However, in graded gratings a groove-to-groove separation of between 300 nm to 900 nm yields an impedance matched graded grating structure as a result of improved adiabatic mode transformation between adjacent grooves, thus enhancing multiwavelength light localization. This optical adiabaticity among the grooves is most notably established when non-resonant modes of each groove are only transferred to the resonant groove. In such an impedance matched graded grating structure, the resonant wavelength is simply determined by examining the resonant wavelengths of the corresponding single groove. At these values of $d$, increasing the groove depth introduces additional resonant modes and thus extends multiwavelength light localization into the near and mid infrared, with improved FWHM and Q factor values for higher order modes in the visible region. This investigation shows that width-based graded gratings can provide adiabatic coupling of multiple wavelengths of light to the resonant modes of the graded grating and thus offers multiwavelength field intensification with eleven orders of magnitude relative to the incident field intensity. These structures which can be reproducibly manufactured open the way for a manifold of high-field intensity sensing applications wherein molecular species of interest on a given width-graded nano-grating surface can be probed essentially simultaneously using several wavelengths spanning from the visible into the infrared and thus lead to high-specificity and high-sensitivity detection.

## Methods

We used COMSOL Multiphysics modelling to study the near-field optical response of MIM graded grating structures in gold with air-filled cavities. The structure of the gratings is illustrated in Figure 1.a where $w$ is the width of the smallest groove with groove-index $j = 0$. $\Delta$ is the groove width gradient parameter, where $\Delta = 0$ defines a uniform lamellar grating while $\Delta \neq 0$ defines a graded lamellar grating wherein the width of each groove increases linearly in increments of $\Delta$ in the $\pm x$ directions. Groove index values spanning $-4 \leq j \leq +4$ and $\Delta \neq 0$ thus create a symmetric graded grating structure around $j = 0$. In our simulations, we chose the largest central groove width to be 30 nm, with the gradient parameter $\Delta = 5$ nm, in order to achieve strong evanescent field coupling between the side-walls of the MIM grooves and thus resulting in relatively flat dispersion curves in contrast to a single interface dispersion curve [20]. Further, we omit the use of a substrate in our simulations considering that the thickness of the metal beneath the grooves is 300 nm which is far greater than the skin depth. All simulations were executed spanning the visible and the near infrared wavelengths, 500 nm < $\lambda_{\text{incident}}$ < 6500 nm. A direct solver (MUMPS[†]) was used to solve Maxwell's equations in the frequency domain. The electric field $E = \left(E_x^2 + E_y^2 + E_z^2\right)^{1/2}$ was measured within each groove, and it was squared to obtain the localized electric field intensity. All the simulations were based on a 2D model where the groove length in the $y$ direction (into the page in Figure 1) was assumed to be infinite considering that groove lengths greater than 1 μm do not significantly alter the resonance behaviour of the structure [14] in the near/mid infrared frequency region and even less so in the visible. The refractive index of gold was obtained from references [30, 31]. All modelled structures were illuminated by normally incident (from the top) TM plane waves. A perfectly matched layer boundary condition was used to avoid contributions from any scattered light at the boundary of the modelled structure. For the case of uniform lamellar grating with an infinite number of grooves, where the repeating unit is one single groove, we used the Floquet periodic boundary condition on both sides of the unit groove. This periodic boundary condition accounted for diffracted orders of the incident light for grating periods larger than the incident light wavelength.

---

[†] MUlit-frontal Massively Parallel Stairs direct solver.

## Acknowledgements

We gratefully acknowledge the support of the Natural Sciences and Engineering Research Council (NSERC) of Canada through the Discovery Grant and Postgraduate Scholarship, the Ontario Research Foundation – Research Excellence program, the Connaught Global Challenge Fund at the University of Toronto (UofT), EMH (Engineering Medicine Health) Seed program at UofT, the Department of Electrical & Computer Engineering at UofT, the Walter Sumner Foundation, and CMC Microsystems.

## Conflict of Interest

The authors declare that there is no conflict of interest.

## Contributions

M. Shayeganniga carried out the theoretical analyses and COMSOL computational modeling, and authored the first manuscript. A. Montazeri contributed to the analyses of the results and reviewed the manuscript, K. Dixon carried out the adiabatic theory development, R. Prinja reviewed the manuscript with an emphasis on its introduction, and N. Kazemi-Zanjani reviewed the manuscript. N. P. Kherani supervised the research and reviewed the article and analyses.

## References

[1] P. Roelli, C. Galland, N. Piro and T. J. Kippenberg, "Molecular cavity optomechanics as a theory of plasmon-enhanced Raman scattering," *Nature nanotechnology,* vol. 11, pp. 164-169, 2015.

[2] N. Kazemi-Zanjani, M. Shayegannia, R. Prinja, A. O. Montazeri, A. Mohammadzadeh, K. Dixon, S. Zhu, P. R. Selvaganapathy, A. Zavodni, N. Matsuura and N. P. Kherani, "Multiwavelength surface-enhanced Raman spectroscopy using rainbow trapping in width-graded plasmonic gratings," *Advanced Optical Materials,* vol. 6, no. 4, pp. 1-8, 2018.

[3] J. Luan, J. J. Morrissey, Z. Wang, H. Gholami Derami, K. K. Liu, S. Cao, Q. Jiang, C. Wang, E. D. Kharasch, R. R. Naik and S. Singamaneni, "Add-on plasmonic path as a universal fluorescence enhancer," *Light: Science & Applications,* vol. 7, no. 29, pp. 1-13, 2018.

[4] B. Chen, H. Zheng, M. Riehn, S. Bok, K. Gangopadhyay, J. Farland, S. Gangopadhyay and M. R. Maschmann, "Enhanced fluorescenec for in situ temperature mapping of photothermally heated aluminum nanoparticles enabled by a plasmonic grating substrate," *Nanotechnology,* vol. 29, pp. 1-10, 2018.

[5] C. Han, M. Lee, S. Callard, C. Seassal and H. Jeon, "Lasing at topoligical edge states in a photonic crystal L3 nanocavity dimer array," *Light: Science & Applications,* vol. 8, no. 40, pp. 1-10, 2019.

[6] M. Najiminaini, F. Vasefi, B. Kaminska and a. J. J. L. Carson, "A three-dimentional plasmonic nanostructure with extraordinary optical transmission," *Journal of Plasmonics,* vol. 8, no. 2, pp. 217-224, 2013.

[7] L. Jiang, T. Yin, A. M. Dubrovkin, Z. Dong, Y. Chen, W. Chen, J. K. W. Yang and Z. Shen, "In-place coherent control of plasmon resonances for plasmonic switching and encoding," *Light: Science & Applications,* vol. 8, no. 21, pp. 1-10, 2019.

[8] Y. Nishijima, L. Rosa and a. S. Juodkazis, "Surface plasmon resonances in periodic and random patterns of gold nano-disks for broadband light harvesting," *Optics Express,* vol. 20, no. 10, pp. 11466-11477, 2012.

[9] A. O. Montazeri, Y. S. Fang, P. Sarrafi and N. P. Kherani, "Rainbow-trapping by adiabatic tuning of intragroove plasmon coupling," *Optics Express,* 2016.

[10] A. Polyakov, M. Zolotorev, P. J. Schuck and H. A. Padmore, "Collective behavior of impedance matched plasmonic nanocavities," *Optics Express,* vol. 20, no. 7, pp. 7685-7693, 2012.

[11] L. Zhu, D. Zhang, X. Wang, Y. Chen, D. Qiu, P. Wang and a. H. Ming, "Dynamically Generating a Large-Area Confined Optical Field with Subwavelength Feature Size," *Optical Society of America,* vol. 53, no. 26, pp. 6091-6095, 2014.

[12] L. Verslegers, P. B. Catrysse, Z. Yu, J. S. White, E. S. Barnard, M. L. Brongersma and a. S. Fan, "Planar Lenses Based on Nanoscale Slit Arrays in a Metallic Flim," *American Chemical Society,* vol. 9, no. 1, pp. 235-238, 2009.

[13] J. M. Manceau, S. Zanotto, I. Sagnes, G. Beaudoin and R. Colombelli, "Optical Critical Coupling into Highly Confining Metal-Insulator-Metal Resonators," *Applied Physics Letter,* vol. 103, p. 091110, 2013.

[14] S. Collin, F. Pardo and a. J.-L. Pelouard, "Waveguiding in nanoscale metallic apertures," *Optics Express,* vol. 15, no. 7, pp. 4310-4320, 2007.

[15] J. L. Perchec, P. Que´merais, A. Barbara. and a. T. Lo´pez-Rı´os., "Why Metallic Surfaces with Grooves a Few Nanometers Deep and Wide May Strongly Absorb Visible Light," *PHYSICAL REVIEW LETTERS,* vol. 100, no. 6, p. 066408, 2008.

[16] Q. Gan and F. J. Bartoli, "Graded Metallic Gratings for Ultrawideband Surface Wave Trapping at THz Frequencies," *IEE Journal of Selected Topics in Quantum Electronics,* vol. 17, no. 1, pp. 102-109, 2011.

[17] D. M. Tanenbaum, C. W. Lo, M. Isaacson and H. G. Craighead, "High resolution electron beam lithography using ZEP-520 and KRS resists at low voltage," *Journal of Vacuum Science & Technology B,* vol. 14, no. 6, pp. 3829-3833, 1996.

[18] L. C. Huang, Z. Wang, J. K. Clark, Y. L. Ho and J. J. Delaunay, "Plasmonic tooth-multilayer structure with high enhancement field for surface enhanced Raman spectroscopy," *Nanotechnology,* vol. 28, no. 12, pp. 1-7, 2017.

[19] A. Yariv and P. Yeh, Photonics-optical electronics in modern communications, New York: Oxford University Press, 2007.

[20] S. A. Maier, PLASMONICS: FUNDAMENTALS AND APPLICATIONS, Springer, 2007.

[21] A. Astilean, P. Lalanne and M. Palamaru, "Light transmission through metallic channels much smaller that the wavelength," *Optics Communications,* vol. 17, pp. 265-273, 2000.

[22] Z. Han, E. Forsberg and S. He, "Surface plasmon bragg gratings formed in metal-insulator-metal waveguides," *IEEE Photonics Technology Letters,* vol. 19, no. 2, pp. 91-93, 2007.

[23] H. T. Miyazaki and a. Y. Kurokawa, "Squeezing visible light waves into a 3-nm thick and 55-nm long plamson cavity," *Physical Review Letters,* vol. 96, no. 9, 2006.

[24] D. J. Griffiths, Introduction to Quantum Mechanics, 2 ed., New Jersey: Pearson Education Inc., 2005.

[25] L. D. Landau and E. M. Lifshitz, Quantum mechanics - Non-relativistic theory, London: Pergamon Press, 1962.

[26] D. d. Ceglia, M. A. Vincenti, M. Scalora and N. A. a. M. J. Bloemer, "Plasmonic band edge effects on the transmission properties of metal gratings," *AIP Advances,* vol. 1, p. 032151, 2011.

[27] F. Neubrech, C. Huck, K. Weber, A. Pucci and H. Giessen, "Surface-enhanced infrared spectroscopy using resonant nanoantennas," *Chemical Reviews,* vol. 117, pp. 5110-5145, 2017.

[28] X. Yang, Z. Sun, T. Low, H. Hu, X. Guo, F. J. Garcia de Abajo, P. Avouris and Q. Dai, "Nanomaterial-based plasmon enhanced infrared spectroscopy," *Advanced Materials,* vol. 30, pp. 1-30, 2018.

[29] A. Polyakov, H. A. Padmore, X. Liang, S. Dhuey, B. Harteneck, J. P. Schuck and S. Cabrini, "Light trapping in plasmonic nanocavities on metal surfaces," *Journal of Vacuum Science & Technology B,* vol. 29, no. 6, pp. 1-4, 2011.

[30] R. W. Christy and P. B. Johnson, "Optical constants of the Noble Metals," *Physical Review B,* vol. 6, no. 12, pp. 4370-4379, 1972.

[31] A. D. Rakic, A. B. Djurisic, J. M. Elazar and M. L. Majewski, "Optical properties of metallic films for vertical-cavity optoelectronic devices," *Applied Optics,* vol. 37, no. 22, pp. 5271-5283, 1998.

## Figures

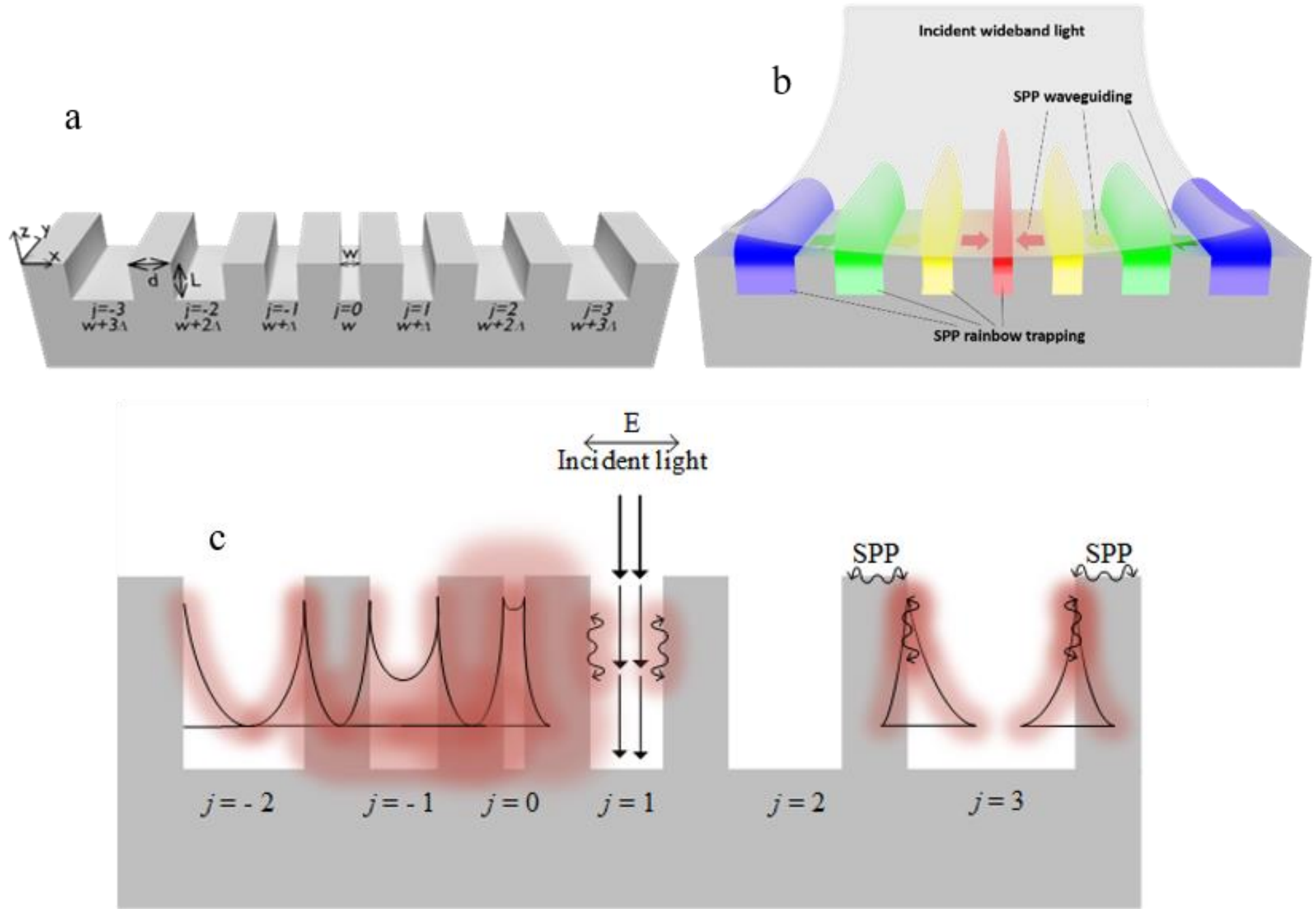

**Figure 1**: a) Schematic of an MIM-based grating with *W*, *d*, *L* and *Δ* representing groove width, groove-to-groove separation, groove depth and the width gradient, respectively; *j* is the groove number. This structure represents a width-graded nano-grating which is symmetric around the centrally situated groove *j* = 0. b) Schematic diagram of rainbow trapping in the plasmonic width-graded nano-grating; smaller the groove width, larger the wavelength of the localized field. Also, effect of SPPs waveguiding on top of the grating in the direction of increased effective index atop the grating is illustrated by the arrows on the surface, and the increasing intensity of the localized field at smaller groove widths. Gradient in the trapped field intensity shows the location of hotspots within and on top of the grooves. c) Illustrating various coupling mechanisms for mode localization (highlighted in red). For brevity, every groove showcases a potential mode coupling mechanism which in general can occur in any other groove. Grooves denoted as *j* = -2 and *j* = 0 show the weak and strong

coupling between the SPPs excited on the opposing side-walls of the groove, respectively. The evanescent mode coupling from adjacent grooves is shown for groove $j$ = -1, specifically coupling with adjoining grooves $j$ = -2 and $j$ = 0. Groove $j$ = 1 shows the coupling of incident light with the cavity mode. Groove $j$ = 3 shows coupling of surface waves with the cavity modes and thus exciting SPPs on the groove's inner walls.

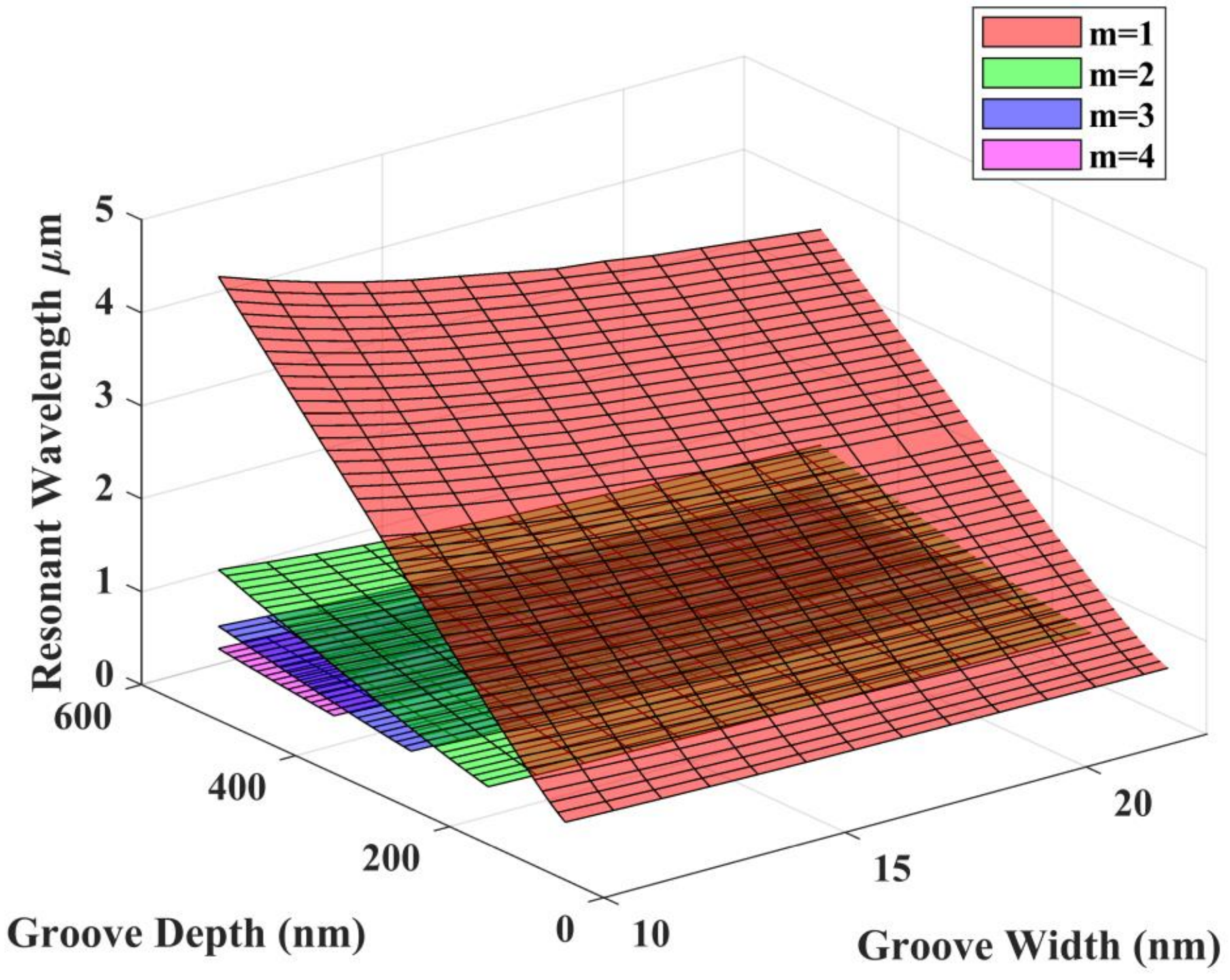

**Figure 2:** Theoretical relationship between groove depth, groove width, and the resonant wavelength of a single groove. In this figure, $m$ corresponds to the mode number. Note that the resonant wavelength can be controlled by changing either the groove width or depth.

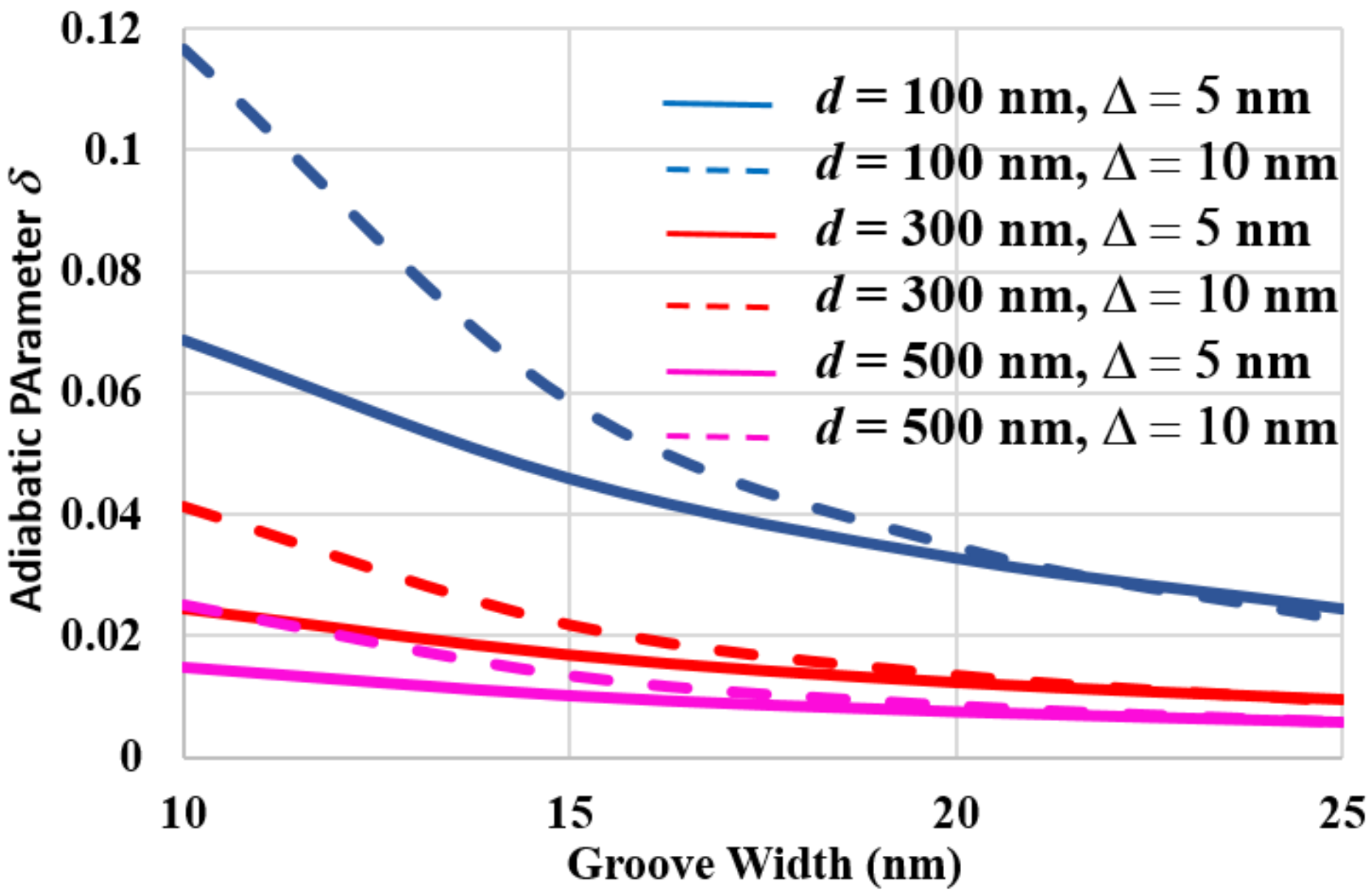

**Figure 3**: Theoretical relationship between the adiabatic parameter ($\delta$) and groove width ($w$) as a function of groove-to-groove separation ($d$), and gradient in groove width ($\Delta$). Note that higher values of $d$ and lower values of $\Delta$ lead to smaller values of $\delta$. The smaller the $\delta$, the better the adiabatic mode transformation between the nano-grooves.

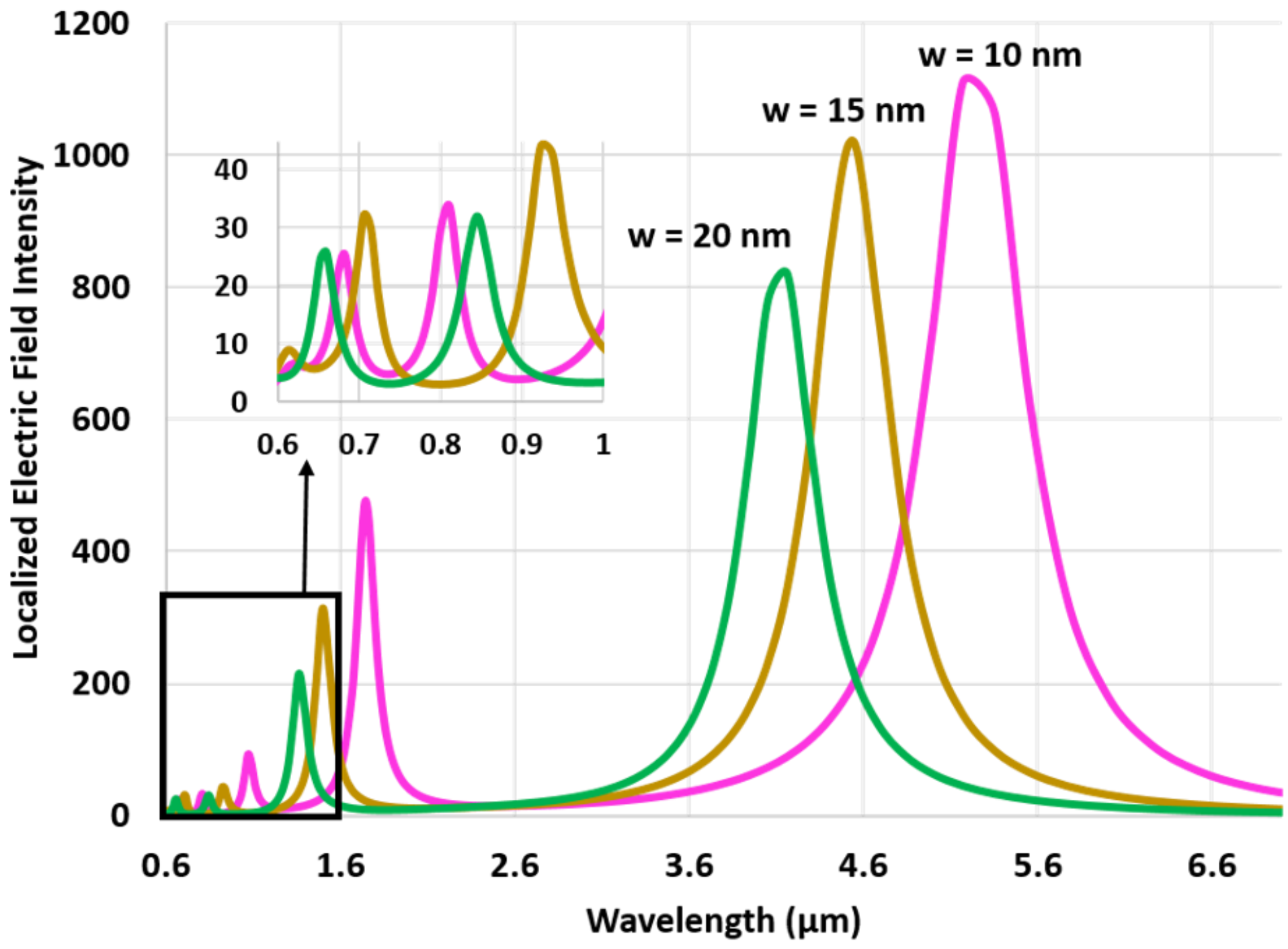

**Figure 4**: Resonance spectrum of a single groove for different groove width *w* and groove depth (*L*) = 500 nm. Wavelength and localized intensity in the groove cavity increases as *w* decreases.

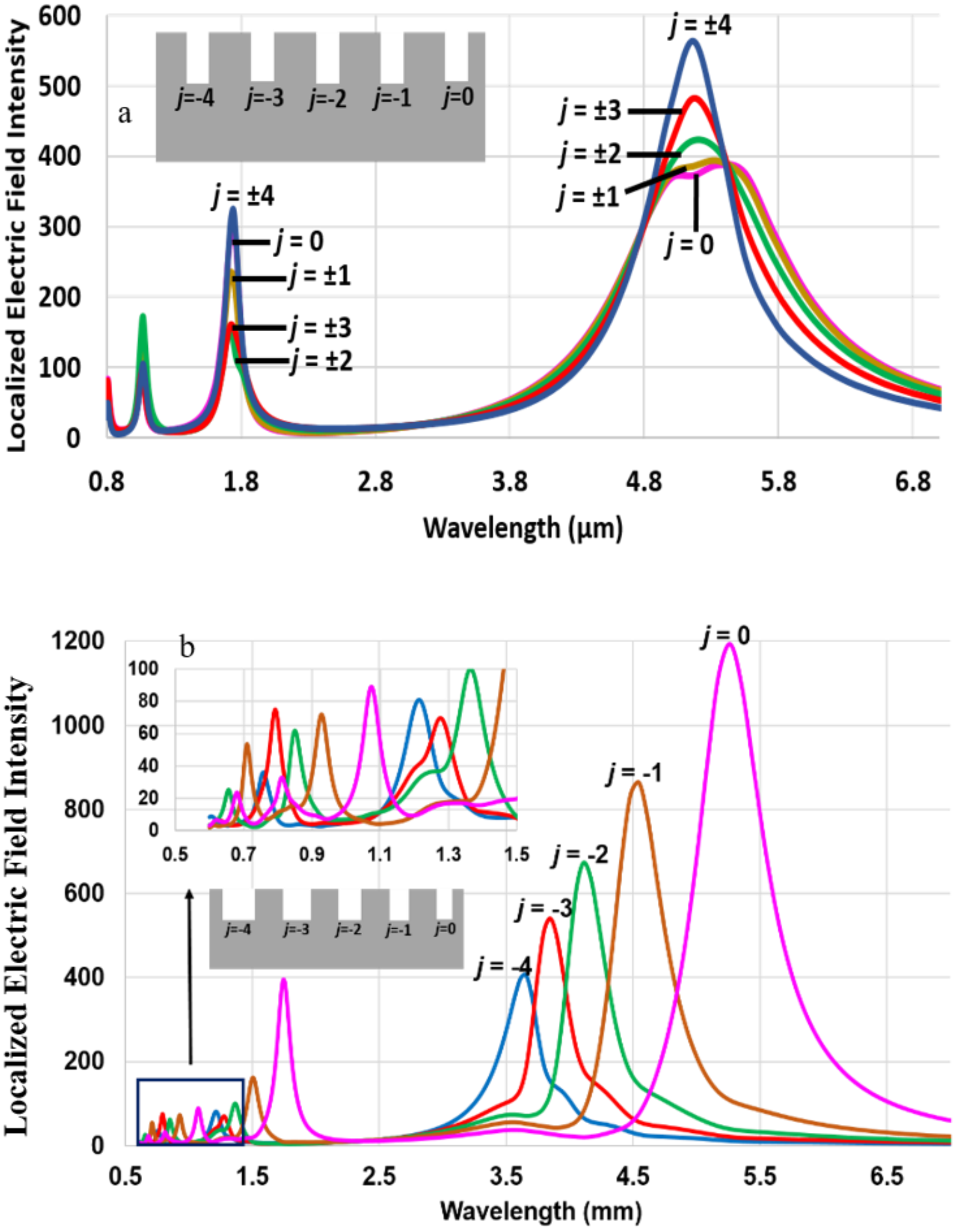

**Figure 5**: a) Localized electric field intensity for a uniform grating with $L$ = 500 nm, $w$ = 10 nm, $n$ = 9 and -4 ≤ $j$ ≤ +4. All the nano-grooves localize identical plasmonic modes, except near the first excited mode around 5 µm where small variations in the resonance mode of different grooves is noted. These variations are related to Eq. (5) where the plasmonic propagation constant obtains two values as a result of mode

propagation in ±$x$ directions. b) Localized electric field intensity for a non-symmetric graded grating with $L$ = 500 nm, $w$ =10 nm, $\Delta$ = 5 nm, $n$ = 5 and $-4 \leq j \leq 0$. Every groove resonates at a particular wavelength of light, and impedance matched graded grating exhibits multiwavelength mode localization in a wide bandwidth of 600 nm to 6 µm. This multiwavelength localization is the result of adiabatic mode transformation.

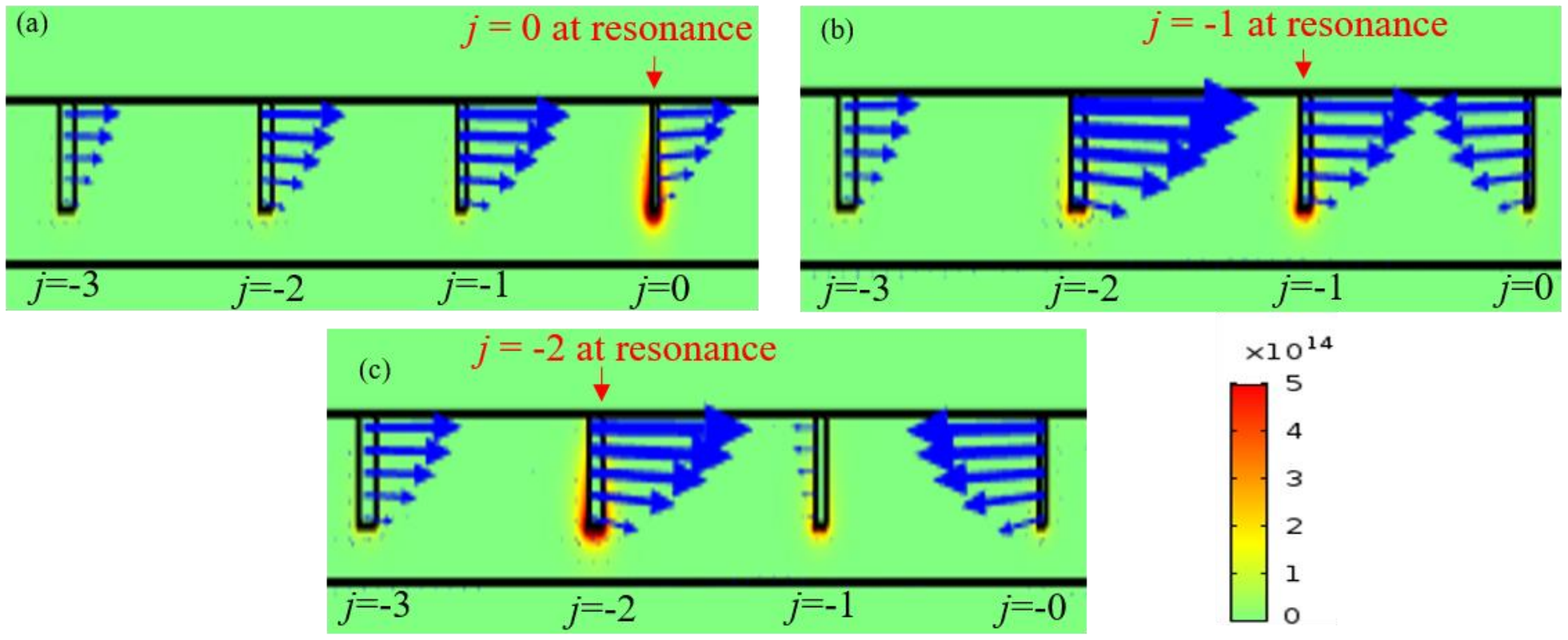

**Figure 6**: (a) $j$ = 0, (b) $j$ = -1, and (c) $j$ = -2 correspond to $\lambda_{\text{resonant}}$ = 5.264 µm, 4.478 µm, and 4.167 µm, respectively, where the groove under resonance is labelled. Total power dissipation density (the rainbow colors surrounding each groove) show the amount of dissipated electric field which is highest at the bottom of the groove under resonance. The blue vectors show the scattered electric field $\vec{E}_x$ within the adjacent grooves. These vectors in any groove point towards the groove under resonance and thus demonstrate how adiabatic mode transformation dictates that all non-resonating grooves direct their scattered field $\vec{E}_x$ towards the resonating groove (see Visualization 1).

**Table 1**: Resonance wavelength, intensity, FWHM, and effective *Q* factor for uniform lamellar gratings with $w$ = 15 nm, $\Delta = 0$, $L$ = 100 nm, $n$> 1

| ***n*** | **2** | | **3** | | | | | | **5** | | | **∞** | | |
|---|---|---|---|---|---|---|---|---|---|---|---|---|---|---|
| ***j*** | 0, 1 | | -1, 1 | 0 | -1, 1 | 0 | -1, 1 | 0 | -2, 2 | -1, 1 | 0 | - | - | - |
| ***d* (nm)** | 5 | 300 | 300 | | 700 | | 1000 | | 300 | | | 5 | 300 | 700 |
| **$\lambda_{resonant}$ (nm)** | 1016 | 1070 | 1153 | 1070 | 1110 | | 1166 | | 1131 | 1070 | 1131<br>1070 | 576 | 1070 | (723)<br>1111 |
| **Intensity** | 65 | 224 | 501 | 523 | 925 | 1946 | 620 | 603 | 311 | 441 | 695<br>691 | 1.81 | 491 | (445)<br>1307 |
| **FWHM (nm)** | 227 | 182 | 158 | 128 | 80 | 80 | 86 | 84 | 145 | 112 | 170 | 237 | 129 | (1)<br>77 |
| **Effective *Q* factor** | 4.5 | 6 | 7.2 | | 14 | | 13.88 | | 8 | | | 2.7 | 8.3 | (723)<br>14.3 |

**Table 2**: Resonance wavelength, intensity, FWHM, and effective *Q* factor for a non-symmetric graded grating with *w* = 10 nm, *Δ* = 5 nm,, *d* = 300 nm, *n* = 5

| ***J*** | -4 | -3 | -2 | -1 | 0 | -4 | -3 | -2 | -1 | 0 |
|---|---|---|---|---|---|---|---|---|---|---|
| ***L* (nm)** | 100 | | | | | 500 | | | | |
| **$\lambda_{resonant}$ (μm)** | | | | | | | | | | |
| 1st mode | 0.92 | 0.952 | 1.07 | 1.13 | 1.25 | 3.66 | 3.85 | 4.11 | 4.55 | 5.26 |
| 2nd mode | | | | | | 1.21 | 1.28 | 1.36 | 1.51 | 1.74 |
| 3rd mode | | | | | | 0.75 | 0.79 | 0.85 | 0.93 | 1.08 |
| 4th mode | | | | | | | | 0.65 | 0.71 | 0.811 |
| 5th mode | | | | | | | | | | 0.68 |
| **Intensity ×$10^{11}$ ($V^2/m^2$)** | | | | | | | | | | |
| 1st mode | 0.28 | 0.67 | 0.88 | 0.99 | 1.17 | 1.13 | 1.51 | 1.89 | 2.42 | 3.34 |
| 2nd mode | | | | | | 0.226 | 0.195 | 0.28 | 0.45 | 1.11 |
| 3rd mode | | | | | | 0.1 | 0.21 | 0.17 | 0.2 | 0.25 |
| 4th mode | | | | | | | | 0.07 | 0.15 | 0.09 |
| 5th mode | | | | | | | | | | 0.06 |
| **FWHM (nm)** | | | | | | | | | | |
| 1st mode | 146 | 165 | 183 | 185 | 153 | 380 | 325 | 343 | 477 | 649 |
| 2nd mode | | | | | | 86.3 | 139 | 105 | 115 | 130 |
| 3rd mode | | | | | | 43 | 49.1 | 55.2 | 54.5 | 65.1 |
| 4th mode | | | | | | | | 34.3 | 32.5 | 60.4 |
| 5th mode | | | | | | | | | | 33.9 |
| **Q factor** | | | | | | | | | | |
| 1st mode | 6.3 | 5.73 | 5.28 | 6 | 8.57 | 9.32 | 11.9 | 12.2 | 9.57 | 8.14 |
| 2nd mode | | | | | | 14 | 8.89 | 12.9 | 13.1 | 13.4 |
| 3rd mode | | | | | | 17.4 | 16.1 | 15.6 | 17 | 16.5 |
| 4th mode | | | | | | | | 19.1 | 21.9 | 13.7 |
| 5th mode ***Q* factor** | | | | | | | | | | 20 |